# Design of Generic Framework for Botnet Detection in Network Forensics

Sukhdilpreet Kaur
Computer Science and Engineering
Punjabi University Regional Centre for Information
Technology and Management
Mohali, India
Sukhdilpreet83@gmail.com

Amandeep Verma
Computer Science and Engineering
Punjabi University Regional Centre for Information
Technology and Management
Mohali, India
vaman71@gmail.com

*Abstract*— With the raise in practice of Internet, in social, personal, commercial and other aspects of life, the cybercrime is as well escalating at an alarming rate. Such usage of Internet in diversified areas also augmented the illegal activities, which in turn, bids many network attacks and threats. Network forensics is used to detect the network attacks. This can be viewed as the extension of network security. It is the technology, which detects and also suggests prevention of the various network attacks. Botnet is one of the most common attacks and is regarded as a network of hacked computers. It captures the network packet, store it and then analyze and correlate to find the source of attack. Various methods based on this approach for botnet detection are in literature, but a generalized method is lacking. So, there is a requirement to design a generic framework that can be used by any botnet detection. This framework is of use for researchers, in the development of their own method of botnet detection, by means of providing methodology and guidelines. In this paper, various prevalent methods of botnet detection are studied, commonalities among them are established and then a generalized model for the detection of botnet is proposed. The proposed framework is described as UML diagrams.

## I. INTRODUCTION

Cyber crime is a huge problem these days. In past few years many researchers have done research on network forensics to reduce the cyber crime. Network forensics is the forensic science that investigates the network traffic and analyzes it for the detection of network attacks. It also tries to find out the source of attack [1]. Botnet is one of the network attacks. It is a network of infected machines called Zombies that have their own life cycle. A controller called botmaster controls Botnets. There is a need to detect the attacks and to prevent them. Detection methods detect and prevent these attacks and try to find out the source of attacks. Many methods of botnet detection are available in literature that are broadly classified into two categories Honeynet based [2] and Passive network traffic monitoring based [3]. Passive network traffic monitoring methods include Botnet Detection Through Fine Flow Classification [4], Detecting Botnets Through Log Correlation [5], Detecting Botnets with Tight Command and Control [6], Botnet Detection by Monitoring Similar Communication Patterns [7], DNS based [8], Data mining based, anomaly based and signature based [9]. All the methods have their specific framework but the generic framework is missing.

In the present study, the focus is around the design of generalized model for botnet detection method because many botnet detection methods are available in literature but there is no such generalized approach. The generic framework of botnet detection is lacking in the literature, which motivates the present study to design a generalized model for botnet detection. This work is indented for those researchers who want to implement a new model for the botnet detection that considers the general architecture.

This paper is organized as follows: Section II presents the background knowledge that describes forensics, network forensics, botnets and botnet detection methods and it also includes the proposed taxonomy of botnet detection methods. The literature review is discussed in section III. The proposal of the generic framework of botnet detection methods is presented in section IV. Future work is stated in section V.

## II. BACKGROUND

### A. Forensics

Forensics is the investigation technique that is used to gather evidences of some criminal activities. Forensic sciences have many branches and network forensics is one of them.

### B. Network Forensics

Network forensics is a branch of forensics science and is the extension of network security. Network security simply detects and prevents the attacks but the network forensics has the capability to do investigation [10]. Network forensics is the investigation technology, which captures the network packets, record them for investigation and then analyze and correlate the recorded network data to find out the source of attacks [1].

### C. Network attacks

With the increase usage of Internet, there is also a rapid increase in cyber crime, which includes various network attacks. Network attacks exploit the vulnerabilities of the system and gain unauthorized access to the system [11]. One of the network attacks is botnet.

*D. Botnets*

Botnet is one of the common network attacks these days. Botnet is defined as a network or group of compromised computers called zombies, which are controlled by a botmaster automatically [3]. The botmaster controls the whole botnet using Command and Control servers [9].

*E. Botnet detection methods*

Botnet detection methods detect the botnet attacks. The botnet detection methods are broadly classified into two categories Honeynet based botnet detection and passive network traffic monitoring [12]. Honeynet is made of collection of more than one honeypot and a honeywall. A honeypot is a system designed to attract the attackers so as to observe their activities and find out solutions and honeywall is software used to do it [2]. Passive network traffic monitoring methods are further classified into IDS based detection, DNS based and data mining based detection techniques.

Figure 1 presents the proposed taxonomy based on the review of literature.

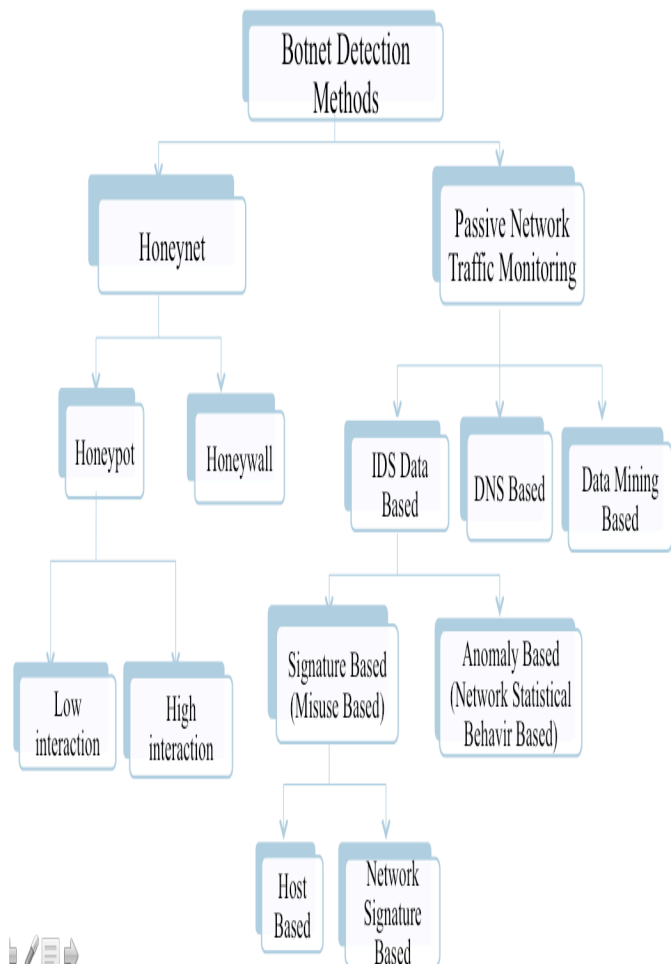

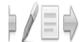
FIGURE 1. Proposed taxonomy of botnet detection methods

III. LITERATURE REVIEW

Cyber crime is a huge problem these days. In past few years many researchers have done research on Network forensics to reduce the cyber crime.

*A. Network Forensics in literature*

Ahmad Almulhem and Issa Traore [1] explored the topic of network forensics and proposed architecture of network forensics system. The proposed architecture manages to collect attack data at network and hosts. It is a capable of bypassing encryption if used by a hacker.

The challenges in deploying a network forensics infrastructure are highlighted by Ahmad Almulhem [10] in "Network Forensics: Notions and Challenges". The various aspects of network forensics and related technologies were presented with limitations of those technologies.

*B. Botnet and Botnet Detection in literature*

J.S.Bhatia, et al [2] presented the introduction to various Internet attacks. They discuss the botnet attacks and propose an approach to detect the botnet attacks that use the IRC and HTTP protocols. The proposed approach is based on Virtual Honeynet based system. They evaluated the approach using real world network traces.

Maryam Feily, et al [3] presented a survey on botnet and botnet detection. The presented survey clarifies what is botnet and also discusses the various botnet detection techniques. Their survey divides the botnet detection techniques into four categories: DNS-based, signature based, anomaly based and mining-based. It also compares the various botnet detection techniques.

Xiaonan Zang, et al. [4] conducted an experiment to observe the discriminating capabilities of the Hierarchical and K mean clustering algorithms and exploring a RTT adjustment procedure to mix the botnet trace with the background Internet traffic. Their experiment has shown the proposed capabilities.

Yousof Al-Hammadi and Uwe Aickelin [5] proposed a new technique to detect the presence of botnets. They used an interception technique to monitor Windows Application Programming Interface *(API)* functions calls made by communication applications and store these calls with their arguments in log files. They proposed an algorithm to detect botnets based on monitoring abnormal activity by correlating the changes in log file sizes from different hosts [5].

Systems detect botnets by examining traffic content for IRC commands or by setting up honeynets. W. Timothy Strayer, et al. [6] proposed an approach for detecting botnets by examining the flow characteristics such as duration, bandwidth, and packet timing that looks for evidence of botnet Command and Control activity. They constructed an architecture that first eliminates traffic that is unlikely to be a part of a network of bots; the remaining traffic is classified into a group that is likely to be part of a botnet, and then correlates the likely traffic to find common communications patterns that would suggest the activity of a botnet. The main focus of this method is on reduction of data set by feeding the traffic packet traces into a series of quick reduction filters.

Hossein Rouhani Zeidanloo and Azizah Bt Abdul Manaf [7] provide taxonomy of botnets C&C channels and they also evaluate the well-known protocols that are being used. They also proposed a general detection framework that focuses on botnets based on P2P and IRC protocols. Their proposed botnet detection framework does not need any prior knowledge like signatures of the botnets.

Sandeep Yadav and A.L. Narasimha Reddy [8] explored the techniques that may utilize the failed domain queries. They present the DNS based botnet detection method.

Yousof Ali Abdulla Al-Hammadi [9] presented an approach that is host-based behavior for the detection of botnets. He monitor the function calls within a time window using various correlation algorithms. He uses an intelligent algorithm that is inspired from the immune systems.

The concepts of network attacks and network security along with cryptography are discussed in [11] by William Stallings.

Alexander V. Barsamian [12] proposed a framework to characterize the network behavior. He starts the research by collecting the network traffic from packet series and hypothesizes that they will characterize the behavior of traffic from threat data. He develops a method to measure the conformity and also detect behavioral changes and also evaluate it. He uses the Kullback-Leibler divergence method for this. He also describe various methods based on K-means approximation for detecting synchronous behavior .He analyze an application of their proposed methods and detect the hosts on the network for the presence of botnet infection.

Robert F. Erbacher, et al [13] introduced a multi-layered architecture to detect the various botnets. They use multiple techniques to detect the old as well as new botnet attacks that cannot be detected by a single technique. For the detection of well-known old botnet attacks, they use signature type techniques and for new botnets, data mining are used.

## IV. PROPOSED FRAMEWORK FOR BOTNET DETECTION

Generic framework of the model for botnet detection is proposed in this section. The proposed framework is composed of some components as described below. The design of the proposed model is given in this paper. To design the proposed model some UML diagrams like class diagrams and activity diagram are used.

### A. Common components of the generalized botnet detection methods

There are some common components that were used by the detection methods that are prevalent in the literature [2,6,7, 13]. This research extracts all the common components followed by the prevalent methods and use the extracted components for designing a generic framework of the model for botnet detection. The extracted common components that are used to design the generic framework of the model for botnet detection method are:

1. Filters
2. Classifiers
3. Correlator
4. Clusters
5. Analyzer

### B. Design

The Model is designed using the UML diagrams. The class diagrams and activity diagram of UML tools are used in the present study. UML diagrams best represents the model and make it easier to understand the concept. UML helps to visualize the designs so that it can be checked against the requirements. There are many UML diagrams. In the present study various classes are used to build the Ontology of Botnet detection method so class diagrams best represent various classes and their subclasses. The flow between the processes of the classes cannot be shown using class diagrams only. To show the flow of data and interaction between the classes, the Activity diagram is used during design phase.

Eight classes are created to design a generic model of botnet detection. The classes are DataSource class that depicts the various sources of data to be analyzed, TrafficScanner class that represents the data capturing tool, PacketFilter class representing the filter components used in botnet detection methods of the literature, FlowClassificationEngine that depicts the classifier component of the prevalent botnet detection methods, PairwiseCorrelator class representing the correlator component of the detection methods studied, Clustering class represents the clusters component extracted from literature, TopologicalAnalyzer class that shows the Analyzer component of existing botnet detection methods and Result class that will show the details of the report generated at the end.

The description of all the classes used in the proposed design is explained here.

#### 1) Class Diagrams

Class diagrams show the static structure of the system to be designed. They represent the entities that share the common characteristics.

Figure 2 shows the class diagram of the class DataSource. It shows the subclasses of the DataSource class. The subclasses of DataSource class are NetworkTrafficInformation, SystemProcessInformation and FileSystemInformation. NetworkTrafficInformation class has a subclass DNS.

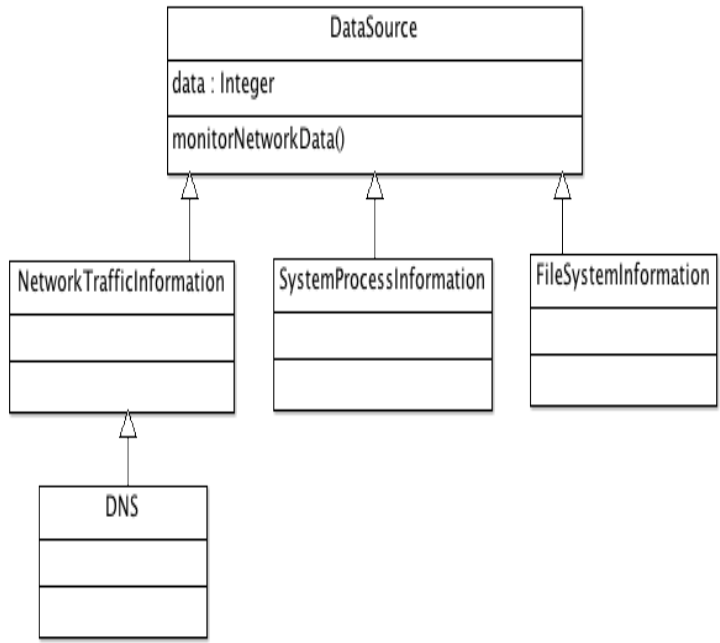

Figure 2. Class diagram of DataSource class

Figure 3 shows the TrafficScanner class. TrafficScanner class has an operation that captures and monitors the data. This class is composed of two classes Agents and Sensors. Agents class gather specific network information and create log files. Sensors class monitors the data in packets and also examine the data. Agents class have two subclasses ActiveAgents and PassiveAgents [14]. ActiveAgents further have a subclass Sniffers that is a data capturing tool. Sensors send the suspicious data to the MarkingModule class.

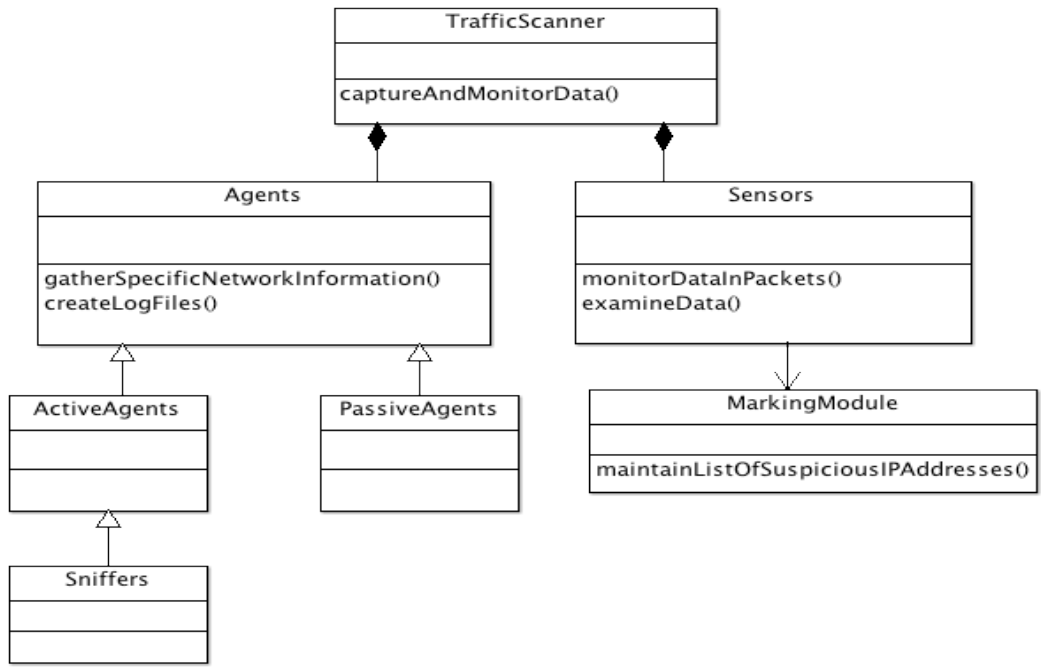

Figure 3. Class diagram of TrafficScanner class

Figure 4 shows the PacketFilter class. This class has attribute called flowAttribute. PacketFilter class has operations detectTrafficContent that detects the contents of the traffic; convertPacketTracesIntoFlowSummaries that convert the packets into flow summaries and eliminate C2 Flows. PacketFilter class is composed of classes QuickDataReduction and IncompleteCommunicationFilter. QuickDataReduction class selects the TCP based flow and the other class that is IncompleteCommunicationFilter class filters out the handshaking process that is SYN-RST exchanges.

DecisionTreeBasedClassifier class is composed of Algorithms class and DataToAdjustAlgorithm class. Algorithm class depicts the algorithms that are used to implement the classifiers and DataToAdjustAlgorithm class represents the data required to adjust the algorithms that are used.

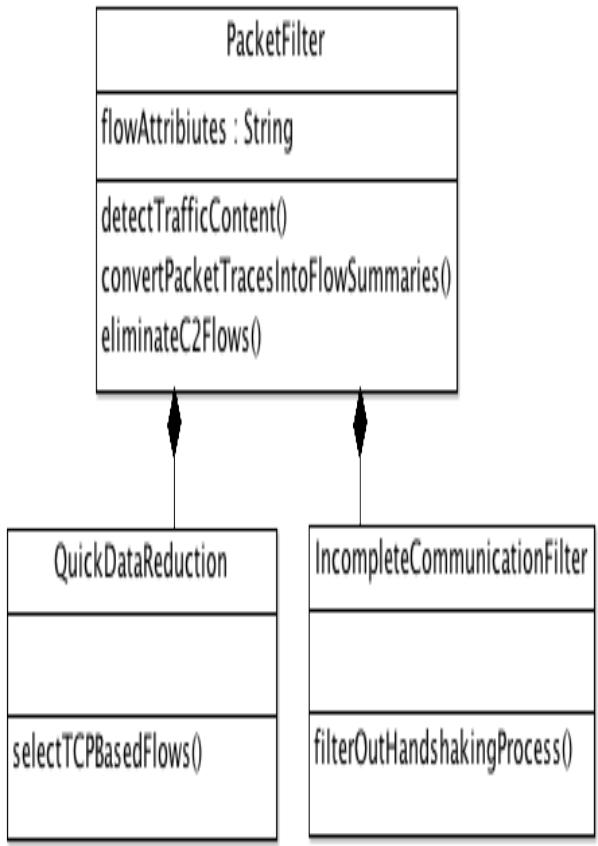

Figure 4. Class diagram of PacketFilter class

Figure 5 describes the FlowClassificationEngine class. It has an attribute payload and two operations classifyTrafficIntoGroups that classifies the incoming traffic into groups and Separate IRCandHTTPtraffic that separates the IRC traffic from the HTTP traffic. FlowClassificationEngine class is composed of FlowBasedDataReduction class and MachineLearningTechniques class. FlowBasedDataReduction class extracts the payload from the flow summaries. MachineLaerningTechniques class does content matching and it has two subclasses SignatureBasedClassifier and DecisionTreeBasedClassifier. SignatureBasedClassifier inspects the payload and DecisionTreeBasedClassifier detect the network anomalies.

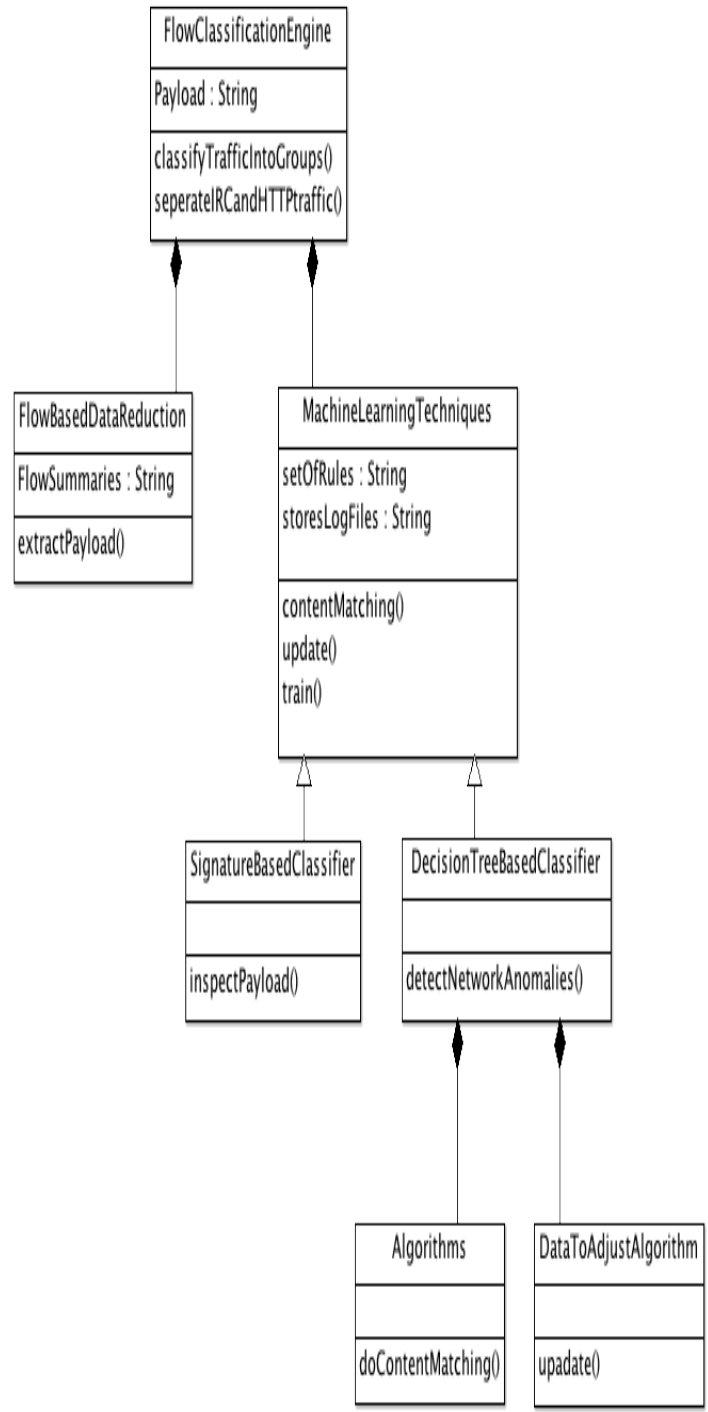

Figure 5. Class diagram of FlowClassificationEngine

Figure 6 shows the PairwiseCorrelator class. It has two attributes flowCharacteristics and payloadCommandSignatures. PairwiseCorrelator class does the pairwise examination of data so as to find out whether one flow is correlated to another flow and then finds the correlation value. It is composed of CorrelationAlgorithm class. CorrelationAlogrithm class implements the correlator.

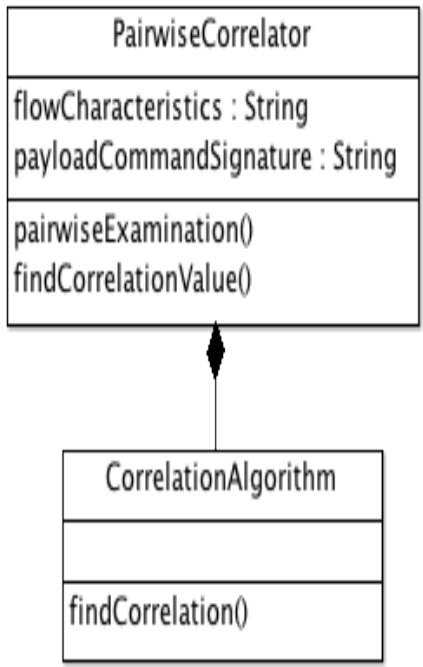

Figure 6. Class diagram of PairwiseCorrelator class

Figure 7 shows the Clustering class. Its attributes are timingCharacteristics and packetSizeCharacteristics. Clustering class group the flows that have similar flow characteristics.

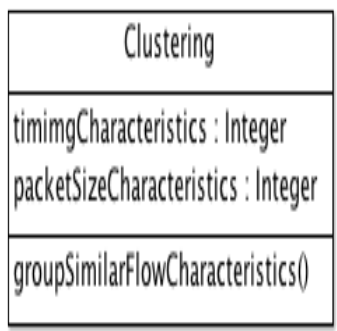

Figure 7. Class diagram of clustering class

Figure 8 shows the TopologicalAnalyzer class, which identifies the controller of the botnets. The controller of botnets is the botmaster. This class finds out the details of the botmaster and sends the details to the Result class.

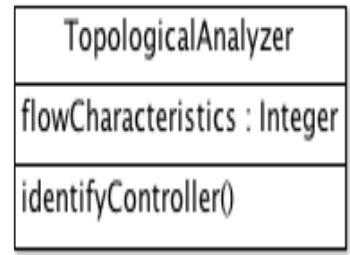

Figure 8. class diagram of TopologicalAnalyzer class

Figure 9 shows Result class. This class represents the result obtained after analyzing the traffic. It shows the details of the controller of the botnets. The details include the IP address of the bot controller along with the name of the bot.

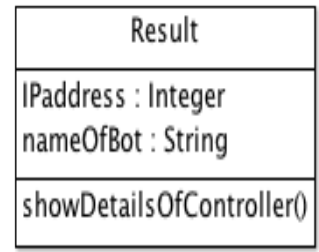

Figure 9. Class diagram of Result class

*2) Activity Diagram*

Activity diagram is also used in this research to design the general model for botnet detection methods. Activity diagrams are best to represent the flow of control between activities and show the system behavior. It is the graphical representation to show that the data moving in the system.

The proposed framework designed for the botnet detection method is demonstrated if Figure 10.

Figure 10 depicts the flow of control between processes of each class in the proposed framework of botnet detection method. Flow starts from the class DataSource that can be network traffic information, system process information and file system information. This class sends the data to the TrafficScanner class, which is a composition of Agents and Sensors. TrafficScanner gathers the specific Network information, create Log files, monitor the data and maintain a list of suspicious IP addresses. It forwards the Packet traces to next class. The next class is Packet Filter, which converts the packet traces to flow summaries, detect traffic content, select the TCP based flow and filters out the handshaking process. Then the remaining flows are sent to the FlowClassificationEngine. FlowClassificationEngine class extract and inspect the payload, does content matching and classify the flow into chat-like and non-chat like flows. The chat-like flows are forwarded to the PairwiseCorrelator that does the pairwise correlation and find the correlated values. The correlated flows are then sent to the Clustering class so that it can group the remaining network traffic with similar flow characteristics and store them in the database. The clusters from the database, in collection, are sent to the TopologicalAnalyzer class that finds out the controller. Controller is the source of attack, which is botmaster. TopologicalAnalyzer class sends the details of the controller to the Result class. Result class presents the result by showing the details of the controller. The details include IP address and name of the botnet.

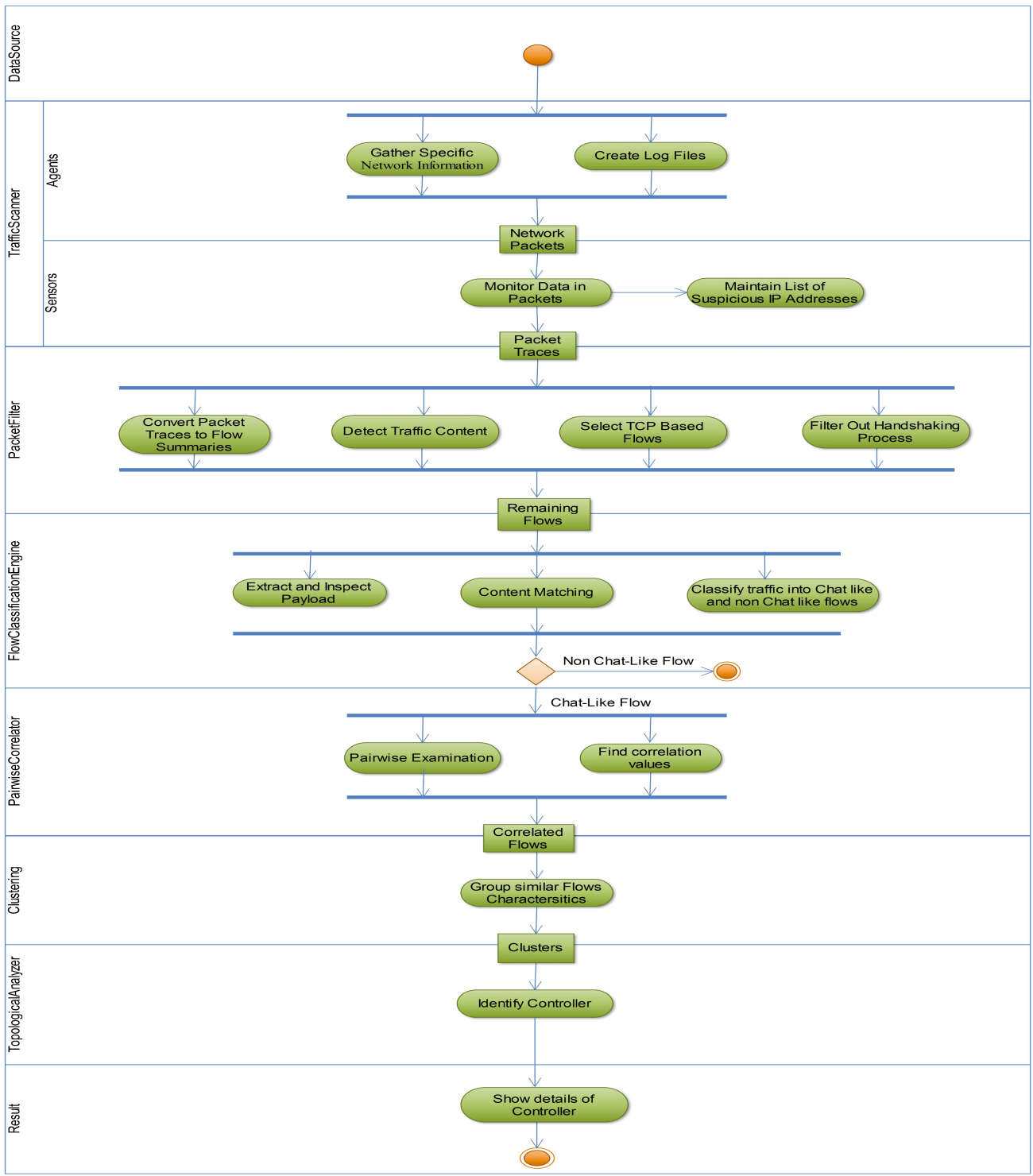

Figure 10. Design of proposed generic framework of botnet detection method

## V. FUTURE WORK

The present study can be extended in future. The further research directions are the generalization with specialization that is to be added to address the specific concerns. A comprehensive version, that can be used to detect attacks, other or in addition to botnet detection, can be devised.

ACKNOWLEDGMENT

We are highly indebted to God for his blessings and love throughout my life and for not letting me down in difficult times. We are grateful to our family for their support.